\documentclass[preprintnumbers, floatfix,letterpaper,aps,prd,epsfig,nofootinbib,
twocolumn
]{revtex4-1}
\usepackage{bm,graphicx,dcolumn,epstopdf,epsf, latexsym,mathbbol, amssymb,amsmath,color,slashed, mathrsfs,mathcomp,simplewick}
\usepackage[center]{subfigure}
\usepackage{multirow}
\usepackage{makecell}
\usepackage[colorlinks,linkcolor=blue,citecolor=blue,urlcolor=blue]{hyperref}

\begin{document}
\allowdisplaybreaks
 \newcommand{\bq}{\begin{equation}}
 \newcommand{\eq}{\end{equation}}
 \newcommand{\bqn}{\begin{eqnarray}}
 \newcommand{\eqn}{\end{eqnarray}}
 \newcommand{\nb}{\nonumber}
 \newcommand{\lb}{\label}
 \newcommand{\f}{\frac}
 \newcommand{\p}{\partial}
\newcommand{\PRL}{Phys. Rev. Lett.}
\newcommand{\PLB}{Phys. Lett. B}
\newcommand{\PRD}{Phys. Rev. D}
\newcommand{\CQG}{Class. Quantum Grav.}
\newcommand{\JCAP}{J. Cosmol. Astropart. Phys.}
\newcommand{\JHEP}{J. High. Energy. Phys.}
\newcommand{\red}{\textcolor{black}}

\title{Waveform of gravitational waves in the ghost-free parity-violating gravities}

\author{Jin Qiao${}^{a, b}$}

\author{Tao Zhu${}^{a, b}$}

\author{Wen Zhao${}^{c, d}$}
\email{Corresponding author: wzhao7@ustc.edu.cn}

 \author{Anzhong Wang${}^{e}$}

\affiliation{${}^{a}$ Institute for Theoretical Physics and Cosmology, Zhejiang University of Technology, Hangzhou, 310032, China\\
${}^b$ United Center for Gravitational Wave Physics (UCGWP), Zhejiang University of Technology, Hangzhou, 310032, China\\
${}^{c}$ CAS Key Laboratory for Research in Galaxies and Cosmology, Department of Astronomy, University of Science and Technology of China, Hefei 230026, China; \\
${}^{d}$ School of Astronomy and Space Sciences, University of Science and Technology of China, Hefei, 230026, China\\
${}^{e}$ GCAP-CASPER, Physics Department, Baylor University, Waco, TX 76798-7316, USA}

\date{\today}

\begin{abstract}

Gravitational waves (GWs) provide an excellent opportunity to test the gravity in the strong gravitational fields. In this article, we calculate the waveform of GWs, produced by the coalescence of compact binaries,
\textcolor{black}{in an extension of the Chern-Simons gravity by including higher derivatives of the coupling scalar field.} By comparing the two circular polarization modes, we find the effects of amplitude birefringence and velocity birefringence of GWs \textcolor{black}{in their propagation} caused by the parity violation in gravity, which are explicitly presented in the GW waveforms by the amplitude and phase modifications respectively.  Combining the two modes, we obtain the GW waveforms in the Fourier domain, and find that the deviations from those in General Relativity are dominated by effects of velocity birefringence of GWs. \textcolor{black}{In addition, we also map the effects of the parity violation on the waveform onto the parameterized post-Einsteinian (PPE) framework and identify explicitly the PPE parameters.}

\end{abstract}


\maketitle

\section{Introduction}
\renewcommand{\theequation}{1.\arabic{equation}} \setcounter{equation}{0}

The discovery of gravitational-wave (GW) compact binary coalescence source GW150914, as well as the other sources, opens the new window of GW astronomy \cite{gw150914,gw170817,gw-other}, which also provides an excellent opportunity to test Einstein's General Relativity (GR) in the strong gravitational fields \cite{test-gr1,test-gr2,test-gr3,test-gr4,gw-nature,gw-white}. In our series of works, we focus on the test of parity symmetry of gravity with GWs. According to Popper¡¯s argument: {\emph{Scientists can never truly ``prove"  that a theory is correct, but rather all we can do is to disprove, or more accurately to constrain a hypothesis. The theory that remains and cannot be disproved by observations becomes the status quo}} \cite{popper}. Therefore, the studies of GW, in particular the calculations of GW waveforms, in the alternative gravitational theories, are the crucial role for the tests of gravity \cite{yunes-review, Ishak:2018his}.

Symmetry permeates nature and is fundamental to all the laws of physics. One example is the parity symmetry, which implies that flipping left and right does not change the laws of physics. As well known, nature is parity violated. Since it was first discovered in weak interactions \cite{Lee-Yang}, the experimental tests become more interesting in the other interactions, including gravity. The birefringence of GWs is a fundamental phenomenon when the parity symmetry is violated in the gravitational sector \cite{Lue:1998mq}. In general, the parity violation can affect the propagation of the GWs in two ways. The first one is it can modify the conventional dispersion relation of the GWs. As a result, the velocities of left-hand and right-hand circular polarization of GWs can be different. This phenomenon is also called {\em velocity birefringence} of GWs. One of examples is the Ho\v{r}ava-Lifshitz theory of gravity \cite{Horava:2009uw} (see refs. \cite{Zhu:2011xe, Zhu:2011yu, Zhu:2012zk, wang2013,zhu2013,Wang:2017brl} for its extensions and a recent review), in which the parity symmetry can be violated by including the third and fifth spatial derivative terms in the action of the theory. 
The second way of the parity violation is that it could change the friction term in the propagation equation of GWs, see examples in the Chern-Simons (CS) modified gravity \cite{Jackiw:2003pm, Lue:1998mq} (see \cite{Alexander:2009tp} for a review). Mainly, such additional friction term will modify the amplitude of GWs, and therefore the amplitude of left-hand circular polarization of gravitational waves will increase (or decrease) during the propagation, while the amplitude for the right-hand modes will decrease (or increase). This phenomenon is also called {\em amplitude birefringence} of GWs, and its corrections to the GWs waveform have been studied in the framework of CS modified gravity in \cite{Alexander:2017jmt, Yagi:2017zhb,Yunes:2010yf,cs1}. 

Recently, based on the specific parity-violating CS modified gravity, a ghost-free parity-violating theory of gravity have been explored in \cite{Crisostomi:2017ugk} by including higher derivatives of the coupling scalar field. The speed of the GW in this theory and its constraints by the GW170817 has been explored in \cite{Nishizawa:2018srh}. A more general modified theory of gravity with parity-violating terms has been constructed in the framework of spatial covariant formulations and its evolution of the GWs in the cosmological background has also been discussed \cite{Gao:2019liu}. In the ghost-free parity-violating gravities, the parity violation can lead to both the velocity and amplitude birefringences in the  propagation of the GWs \cite{zhao2019}. In this paper, we study in detail the effects of both velocity and amplitude birefringences on the GWs waveform. Decomposing the GWs into the left-hand and right-hand circular polarization modes, we find that the effects of velocity and amplitude birefringences can be explicitly presented by the modifications in the GW phase and amplitude respectively. Converting the circular polarizations to the general plus and cross modes, we obtain GW waveforms in the frequency domain, and derive the correction terms in the amplitude and phase of GWs, relative to the corresponding results in GR. The corresponding parameterized post-Einsteinian parameters in the general ghost-free parity-violating gravities are also identified.

This paper is organized as follows. In Secs. \ref{sec2} and \ref{sec3}, we briefly introduce the theories of ghost-free parity-violating gravity, and the propagation of GWs in these theories of gravity respectively. In Sec.~\ref{sec4}, we discuss the amplitude and velocity birefringence effects of GWs. In Sec.~\ref{sec5}, we calculate the waveform of GWs produced by the coalescence of compact binary systems, and particularly focus on the deviations from those in GR. The summary of this work is given in Sec.~\ref{sec6}.

Throughout this paper, the metric convention is chosen as $(-,+,+,+)$, and greek indices $(\mu,\nu,\cdot\cdot\cdot)$ run over $0,1,2,3$ \textcolor{black}{and the latin indices $(i, j, k, \cdots)$ run over $1, 2, 3$.} We set the units to $c=\hbar=1$.

\section{Parity-violating gravities \label{sec2}}
\renewcommand{\theequation}{2.\arabic{equation}} \setcounter{equation}{0}

We consider parity-violating gravity with the action of the form
\bqn\lb{action}
S &=& \frac{1}{16\pi G} \int d^4 x \sqrt{-g}(R+\mathcal{L}_{\rm PV})  \nb\\
&&~~ + \int d^4 x \sqrt{-g} ( \mathcal{L}_{\phi} + \mathcal{L}_{\rm other}),
\eqn
where $R$ is the Ricci scalar, $\mathcal{L}_{\rm PV}$ is a parity-violating Lagrangian, $\mathcal{L}_\phi$ is the Lagrangian for a scalar field, which may be coupled non-minimally to gravity, and $\mathcal{L}_{\rm other}$ denotes other matter fields. As one of the simplest examples, we consider the action of the scalar field
\bqn
\mathcal{L}_\phi =   \frac{1}{2} g^{\mu \nu} \partial_\mu \phi \partial_\nu \phi +V(\phi).
\eqn
Here $V(\phi)$ denotes the potential of the scalar field. The parity-violating Lagrangian $\mathcal{L}_{\rm PV}$ has different expressions for different theories. CS modified gravity with Pontryagin term coupled with a scalar field is a widely studied parity-violating gravity in the previous works. The Lagrangian of CS reads \cite{Alexander:2009tp}
\bqn
\mathcal{L}_{\rm CS} = \frac{1}{8}\vartheta(\phi) \varepsilon^{\mu\nu\rho\sigma} R_{\rho\sigma \alpha\beta} R^{\alpha \beta}_{\;\;\;\; \mu\nu},
\eqn
with $\varepsilon^{\rho \sigma \alpha \beta}$ being the Levi-Civit\'{a} tensor defined in terms of the the antisymmetric symbol $\epsilon^{\rho \sigma \alpha \beta}$ as $\varepsilon^{\rho \sigma \alpha \beta}=\epsilon^{\rho \sigma \alpha \beta}/\sqrt{-g}$ and the CS coupling coefficient $\vartheta(\phi)$ being an arbitrary function of $\phi$. CS modified gravity is an effective extension of GR that captures leading-order, gravitational parity-violating term. The similar versions of this theory were suggested in the context of string theory \cite{string}, and three-dimensional topological massive gravity \cite{massive}. However, this theory has higher-derivative field equation, which induces the dangerous Ostrogradsky ghosts. For this reason, CS modified gravity can only be treated as a low-energy truncation of a fundamental theory. To cure this problem, the extension of CS gravity by considering the terms which involve the derivatives of a scalar field is recently proposed in \cite{Crisostomi:2017ugk}. $\mathcal{L}_{\rm PV1}$ is the Lagrangian containing the first derivative of the scalar field, which is given by
\bqn
\mathcal{L}_{\rm PV1} &=& \sum_{\rm A=1}^4  a_{\rm A}(\phi, \phi^\mu \phi_\mu) L_{\rm A},\label{lv1}\\
L_1 &=& \varepsilon^{\mu\nu\alpha \beta} R_{\alpha \beta \rho \sigma} R_{\mu \nu\; \lambda}^{\; \; \;\rho} \phi^\sigma \phi^\lambda,\nonumber\\
L_2 &=&  \varepsilon^{\mu\nu\alpha \beta} R_{\alpha \beta \rho \sigma} R_{\mu \lambda }^{\; \; \;\rho \sigma} \phi_\nu \phi^\lambda,\nonumber\\
L_3 &=& \varepsilon^{\mu\nu\alpha \beta} R_{\alpha \beta \rho \sigma} R^{\sigma}_{\;\; \nu} \phi^\rho \phi_\mu,\nonumber\\
L_4 &=&  \varepsilon^{\mu\nu\rho\sigma} R_{\rho\sigma \alpha\beta} R^{\alpha \beta}_{\;\;\;\; \mu\nu} \phi^\lambda \phi_\lambda,\nonumber
\eqn
with $\phi^\mu \equiv \nabla^\mu \phi$, and $a_{\rm A}$ are a priori arbitrary functions of $\phi$ and $\phi^\mu \phi_\mu$. In order to avoid the Ostrogradsky modes in the unitary gauge (where the scalar field depends on time only), it is required that $4a_1+2 a_2+a_3 +8 a_4=0$. With this condition, the Lagrangian in Eq.(\ref{lv1}) does not have any higher order time derivative of the metric, but only higher order space derivatives.

One can also consider the terms which contain second derivatives of the scalar field. Focusing on only these that are linear in Riemann tensor and linear/quadratically in the second derivative of $\phi$, the most general Lagrangian $\mathcal{L}_{\rm PV2}$ is given by \cite{Crisostomi:2017ugk}
\bqn
\mathcal{L}_{\rm PV2} &=& \sum_{\rm A=1}^7 b_{\rm A} (\phi,\phi^\lambda \phi_\lambda) M_{\rm A},\\
M_1 &=& \varepsilon^{\mu\nu \alpha \beta} R_{\alpha \beta \rho\sigma} \phi^\rho \phi_\mu \phi^\sigma_\nu,\nonumber\\
M_2 &=& \varepsilon^{\mu\nu \alpha \beta} R_{\alpha \beta \rho\sigma} \phi^\rho_\mu \phi^\sigma_\nu, \nonumber\\
M_3 &=& \varepsilon^{\mu\nu \alpha \beta} R_{\alpha \beta \rho\sigma} \phi^\sigma \phi^\rho_\mu \phi^\lambda_\nu \phi_\lambda, \nonumber\\
M_4 &=& \varepsilon^{\mu\nu \alpha \beta} R_{\alpha \beta \rho\sigma} \phi_\nu \phi_\mu^\rho \phi^\sigma_\lambda \phi^\lambda, \nonumber\\
M_5 &=& \varepsilon^{\mu\nu \alpha \beta} R_{\alpha \rho\sigma \lambda } \phi^\rho \phi_\beta \phi^\sigma_\mu \phi^\lambda_\nu, \nonumber\\
M_6 &=& \varepsilon^{\mu\nu \alpha \beta} R_{\beta \gamma} \phi_\alpha \phi^\gamma_\mu \phi^\lambda_\nu \phi^\lambda, \nonumber\\
M_7 &=& (\nabla^2 \phi) L_1.\nonumber
\eqn
with $\phi^{\sigma}_\nu \equiv \nabla^\sigma \nabla_\nu \phi$. Similarly, in order to avoid the Ostrogradsky modes in the unitary gauge, the following conditions should be imposed: $b_7=0$, $b_6=2(b_4+b_5)$ and $b_2=-A_*^2(b_3-b_4)/2$, where $A_*\equiv \dot{\phi}(t)/N$ and $N$ is the lapse function. In this paper, we consider a general scalar-tensor theory with parity violation, which contains all the terms mentioned above. So, the parity-violating term in Eq.(\ref{action}) is given by
\bqn
\mathcal{L}_{\rm PV} = \mathcal{L}_{\rm CS} + \mathcal{L}_{\rm PV1} + \mathcal{L}_{\rm PV2}.
\eqn
Therefore, the CS modified gravity in \cite{Alexander:2009tp}, and the ghost-free parity-violating gravities discussed in \cite{Crisostomi:2017ugk} are all the particular cases of this Lagrangian. The coefficients $\vartheta$, $a_{\rm A}$ and $b_{\rm A}$ depend on the scalar field $\phi$ and its evolution.

\section{Gravitational waves in parity-violating gravities \label{sec3}}
\renewcommand{\theequation}{3.\arabic{equation}} \setcounter{equation}{0}

Let us investigate the propagation of GW in the theories of gravity with the action given by (\ref{action}). We consider the GWs propagating on a homogeneous and isotropic background. The spatial metric in the flat Friedmann-Robertson-Walker universe is written as
\bqn
g_{ij}=a^2(\tau) (\delta_{ij} + h_{ij}(\tau, x^i)),
\eqn
where $\tau$ denotes the conformal time, which relates to the cosmic time $t$ by $dt=ad\tau$, and $a$ is the scale factor of the universe. Throughout this paper, we set the present scale factor $a_0=1$. $h_{ij}$ is the GW, which represents the transverse and traceless metric perturbations, i.e.,
\bqn
\partial^i h_{ij} =0 = h_i^i.
\eqn
With the above definitions, we need to derive the equation of motion for the GWs. For this purpose, we first need to substitute the metric perturbation into the action (\ref{action}) and expand it to the second order in $h_{ij}$. After tedious calculations, we find
\bqn
S^{(2)} = \frac{1}{16\pi G} \int d\tau d^3 x a^4(\tau) \left[ \mathcal{L}_{\rm GR}^{(2)} + \mathcal{L}_{\rm PV}^{(2)}\right],
\eqn
where
\bqn
\mathcal{L}_{\rm GR}^{(2)} &=& \frac{1}{4 a^2} \left[ (h'_{ij})^2 - (\partial_k h_{ij})^2\right],\\
\mathcal{L}_{\rm PV}^{(2)} &=& \frac{1}{4 a^2} \left[\frac{c_1(\tau)}{a \red{M_{\rm PV}}} \epsilon^{ijk}h_{il}' \partial_j h_{kl}'+ \frac{c_2(\tau)}{a \red{M_{\rm PV}} }   \epsilon^{ijk}\partial^2h_{il} \partial_j h_{kl}\right].\nb\\
\eqn
Here, $\red{M_{\rm PV}}$ labels the parity-violating energy scale in this theory. $c_1$ and $c_2$ are the coefficients normalized by the energy scale $\red{M_{\rm PV}}$, which are given by
\bqn
\frac{ c_1(\tau)}{\red{M_{\rm PV}}} &=& \dot{\vartheta}-4 \dot{a_1}\dot{\phi}^2 -8 a_1\dot{\phi}\ddot{\phi} + 8a_1 H \dot{\phi}^2- 2\dot{a_2}\dot{\phi}^2 - 4a_2\dot{\phi}\ddot{\phi}\nb\\
&&+\dot{a_3}\dot{\phi}^2 +2a_3\dot{\phi}\ddot{\phi} -4a_3 H \dot{\phi}^2-4\dot{a_4}\dot{\phi}^2 -8a_4\dot{\phi}\ddot{\phi}\nb\\
&&-2 b_1\dot{\phi}^3+4b_2\left(2 H\dot{\phi}^2-\dot{\phi}\ddot{\phi}\right)\nb\\
&&+2b_3\left(\dot{\phi}^3\ddot{\phi}- H \dot{\phi}^4\right)+2b_4\left(\dot{\phi}^3\ddot{\phi}- H \dot{\phi}^4\right)\nb\\
&&-2b_5 H \dot{\phi}^4+2b_7\dot{\phi}^3\ddot{\phi},\label{c1-1}\\
\frac{c_2(\tau)}{\red{M_{\rm PV}}} &=&  \dot{\vartheta} - 2\dot{a_2}\dot{\phi}^2 -4a_2\dot{\phi}\ddot{\phi} -\dot{a_3}\dot{\phi}^2 -2a_3\dot{\phi}\ddot{\phi} \nb\\
&&-4\dot{a_4}\dot{\phi}^2 -8a_4\dot{\phi}\ddot{\phi}.\label{c2-2}
\eqn
In this paper, a {\emph{dot}} denotes the derivative with respect to the cosmic time $t$, and $H \equiv \dot{a}/a$ is the Hubble parameter. \red{In Chern-Simons gravity, as shown in \cite{CS_E}, the energy scale $M_{\rm PV}$ of the parity violation has been constrained to be $M_{\rm PV} > 33 \;{\rm meV}$ by using the observation of binary pulsars. One can also constrain $M_{\rm PV}$ by testing the amplitude birefringence effects in the observations of GWs \cite{Yunes:2010yf}. For the ghost-free parity violating gravities, as we shown later in this paper, the parity violation can lead to another birefringence phenomenon, the velocity birefringence effects in GWs. It is expected that one could be able to constrain $M_{\rm PV}$ more tightly by testing velocity birefringence with current and future GW detections.
}

We find that the terms $b_A$ appear only in the $c_1$ coefficient, while the terms $\vartheta$ and $a_A$ appear in both coefficients $c_1$ and $c_2$, which is consistent with the statement in \cite{Nishizawa:2018srh}. From these expressions, we obtain the following quantities,
\bqn
\frac{ c_1(\tau)-c_2(\tau)}{\red{M_{\rm PV}}} &=&-4 \dot{a_1}\dot{\phi}^2 -8 a_1\dot{\phi}\ddot{\phi} + 8a_1 H \dot{\phi}^2\nb\\
&&+2\dot{a_3}\dot{\phi}^2 +4a_3\dot{\phi}\ddot{\phi} -4a_3 H \dot{\phi}^2\nb\\
&&-2 b_1\dot{\phi}^3+4b_2\left(2 H\dot{\phi}^2-\dot{\phi}\ddot{\phi}\right)\nb\\
&&+2b_3\left(\dot{\phi}^3\ddot{\phi}- H \dot{\phi}^4\right)+2b_4\left(\dot{\phi}^3\ddot{\phi}- H \dot{\phi}^4\right)\nb\\
&&-2b_5 H \dot{\phi}^4+2b_7\dot{\phi}^3\ddot{\phi},
\eqn
which will be used in the following discussion.


We consider the GWs propagating in the vacuum, and ignore the source term. Varying the action with respect to $h_{ij}$, we obtain the field equation for $h_{ij}$,
\bqn
&&h_{ij}'' + 2 \mathcal{H} h_{ij}'  - \partial^2 h_{ij}  \nb\\
&&~+ \frac{\epsilon^{ilk}}{a\red{M_{\rm PV}}} \partial_l \Big[ c_1 h_{jk}'' + (\mathcal{H}c_1+c_1') h_{jk}' - c_2 \partial^2 h_{jk}\Big]=0,\nb\\
\eqn
where $\mathcal{H}\equiv a'/a$, and a {\emph{prime}} denotes the derivative with
respect to the conformal time $\tau$.

In the parity-violating gravities, it is convenient to decompose the GWs into the circular polarization modes. To study the evolution of $h_{ij}$, we expand it over spatial Fourier harmonics,
\bqn
h_{ij}(\tau, x^i) = \sum_{A={\rm R, L}} \int \frac{d^3 k}{(2\pi)^3}  h_A(\tau, k^i)e^{i k_i x^i} e_{ij}^{A}(k^i),\nb\\
\eqn
where $e_{ij}^A$ denote the circular polarization tensors and satisfy the relation
\bqn
\epsilon^{i j k} n_i e_{kl}^A = i \rho_A e^{jA}_{~l},
\eqn
with $\rho_{\rm R}=1$ and $\rho_{\rm L} =-1$. We find that the propagation equations of these two modes are decoupled, which can be casted into the form
\bqn\lb{eom}
h_A'' + (2+\nu_A) \mathcal{H} h_A' + (1+\mu_A) k^2 h_A=0,
\eqn
where
\bqn
\nu_A &=& \frac{\rho_A k (c_1 \mathcal{H} -c_1' )/(a \mathcal{ H}  \red{M_{\rm PV}})}{1- \rho_A k c_1/(a \red{M_{\rm PV}})}, \lb{nuA}\\
\mu_A &=& \frac{\rho_A k (c_1 -c_2)/(a \red{M_{\rm PV}})}{1- \rho_A k c_1/(a \red{M_{\rm PV}})}.\lb{muA}
\eqn

The effects of the parity violation terms are fully characterized by two parameters: $\mu_A$ and $\nu_A$. The parameter $\mu_A$ determines the speed of the gravitational waves, which leads to different velocities of left-hand and right-hand circular polarizations of GWs. For the left-hand and right-hand GWs, we find $\mu_A$ have the same value but opposite signs. As a result, the arrival times of the two circular polarization modes could be different. The parameter $\nu_A$, on the other hand, can provide an amplitude modulation to the gravitational waveform, therefore the amplitude of left-hand circular polarization of gravitational waves will increase (or decrease) during the propagation, while the amplitude for the right-hand modes will decrease (or increase). It is interesting to note that in CS modified gravity, since $c_1 = c_2= \red{M_{\rm PV}} \dot \vartheta$, which follows that $\mu_A = 0$. So, there are no modifications on the velocity of GWs and the parity violation can only affect the amplitude. However, in the ghost-free parity-violating gravities with $\mathcal{L}_{\rm PV1}$ and/or $\mathcal{L}_{\rm PV2}$, both terms $\nu_A$ and $\mu_A$ are nonzero. Therefore, both amplitude and velocity birefringence effects exist during the propagation of GWs. In the following section, we shall study these two effects in detail.

\section{Amplitude and velocity birefringences \label{sec4} }
\renewcommand{\theequation}{4.\arabic{equation}} \setcounter{equation}{0}

In this section, we study the phase and amplitude corrections to the waveform of GWs arising from the parameters $\nu_A$ and $\mu_A$. In parity-violating gravities described by (\ref{action}), assuming $k/\red{M_{\rm PV}}\ll 1$, the expressions of $\nu_A$ and $\mu_A$ in Eqs.(\ref{nuA}) and (\ref{muA}) can be written as
\bqn
\mu_A = \rho_A (c_1 -c_2) \left({k}/{a \red{M_{\rm PV}}}\right),\\
\nu_A = \rho_A (c_1 - {c_1'}/{\mathcal{H}}) \left({k}/{a \red{M_{\rm PV}}}\right).
\eqn
\red{The assumption $k \ll M_{\rm PV}$ is based on the consideration that $M_{\rm PV} > 33\; {\rm meV} \sim 10^{13} {\rm Hz}$ \cite{CS_E}, which is about 10 orders of magnitude greater than the sensitive frequency range $f \sim k/2\pi \lesssim 10^3 {\rm Hz}$ for most of GW detectors.}


We further decompose $h_A$ as
\bqn
h_A = \bar h_A e^{-i \theta(\tau)}, \lb{decom} \\
\bar h_A = \mathcal{A}_A e^{- i \Phi(\tau)},\lb{decom1}
\eqn
where $\bar h_A$ satisfies
\bqn\lb{hAbar}
\bar h_A '' + 2 \mathcal{H} \bar h'_A + (1+\mu_A) k^2\bar h_A=0,
\eqn
 here $\mathcal{A}_A$ denotes the amplitude of $\bar h_A$ and $\Phi(\tau)$ is the phase. With the above decomposition,  $\theta (\tau)$ denotes the correction arising from $\nu_A$, while the corrections due to $\mu_A$ is included in $\bar h_A$.

\subsection{Phase modifications}

We first concentrate on the corrections arising from the parameter $\mu_A$, which leads to velocity difference of the two circular polarizations of GWs. To proceed, we define $\bar u_k^{A}(\tau) = \frac{1}{2} a(\tau) M_{\rm Pl}\bar{h}_A (\tau) $ and then Eq. (\ref{hAbar}) can be written as
\bqn
\frac{d^2 \bar u_k^{A}}{d\tau^2} + \left(\omega_A^2 - \frac{a''}{a}\right) \bar u_k^{A}=0,
\eqn
where
\bqn
\omega_A^2(\tau) = k^2(1+\mu_A),
\eqn
is the modified dispersion relation. With this relation, the speed of the graviton reads
\bqn
{v_A^2} = {k^2}/{\omega_A^2} \simeq  1- \rho_A  (c_1-c_2) \left( {k}/{a\red{M_{\rm PV}}}\right),
\eqn
which leads to
\bqn
{v_A} \simeq 1- ({1}/{2}) \rho_A  (c_1-c_2) \left( {k}/{a\red{M_{\rm PV}}}\right).
\eqn
Since $\rho_A$ have the opposite signs for left-hand and right-hand polarization modes, we find that one mode is superluminal and the other is subluminal. Considering a graviton emitted radially at $r=r_e$ and received at $r=0$, we have
\bqn
\frac{dr }{dt} = - \frac{1}{a} \left[1- \frac{1}{2}\rho_A  (c_1-c_2) \left( \frac{k}{a\red{M_{\rm PV}}}\right)\right].
\eqn
Integrating this equation from the emission time ($r=r_e$) to arrival time ($r=0$), one obtains
\bqn
r_e = \int^{t_0}_{t_e} \frac{dt}{a(t)} - \frac{1}{2}\rho_A \left(\frac{k}{ \red{M_{\rm PV}}}\right) \int_{t_e}^{t_0} \frac{c_1-c_2}{a^{2}}dt.
\eqn

Consider gravitons with same $\rho_{A}$ emitted at two different times $t_e$ and $t_e'$, with wave numbers $k$ and $k'$, and received at corresponding arrival times $t_0$ and $t_0'$ ($r_e$ is the same for both). Assuming $\Delta t_e\equiv t_e-t_e'\ll a/\dot{a}$, then the difference of their arrival times is given by
\[
\Delta t_{0}=(1+z)\Delta t_e+\frac{1}{2}\rho_A\frac{k-k'}{\red{M_{\rm PV}}}\int_{t_e}^{t_0} \frac{c_1-c_2} {a^{2}}dt,
\]
where $z\equiv 1/a(t_e)-1$ is the cosmological redshift. 
Let us focus on the GW signal generated by non-spinning, quasi-circular inspiral in the post-Newtonian approximation. Relative to the GW in GR, the term $\mu_{\rm A}$ modifies the phase of GW $\Phi(\tau)$. The Fourier transform of $\bar{h}_{\rm A}$ can be obtained analytically in the stationary phase approximation, where we assume that the phase is changing much more rapidly than the amplitude, which is given by \cite{spa}
\bqn
\tilde{\bar{h}}_{ A}(f)=\frac{{\mathcal{A}}_{ A}(f)}{\sqrt{df/dt}}e^{i\Psi_A(f)},
\eqn
where $f$ is the GW frequency at the detector, and $\Psi$ is the phase of GWs. In \cite{Mirshekari:2011yq}, it is proved that, the difference of arrival times as above induces the modification of GWs phases $\Psi_A$ as follows,
\bqn
\Psi_A(f) = \Psi_A^{\rm GR} (f) + \delta \Psi_A(f),
\eqn
with
\bqn \lb{delta_Phi}
\delta \Psi_A(f) = \xi_A u^{2},
\eqn
where
\bqn
\xi_A &=& \frac{\rho_A}{\red{M_{\rm PV}} \mathcal{M}^{2}} \int_{t_e}^{t_0} {\frac{c_1-c_2}{a^{2}}}dt, \\
u&=& \pi \mathcal{M} f.
\eqn
The quantity $\mathcal{M} = (1+z) \mathcal{M}_{\rm c}$ is the measured chirp mass, and $\mathcal{M}_{\rm c}\equiv (m_1 m_2)^{3/5}/(m_1+m_2)^{1/5}$ is the chirp mass of the binary system with component masses $m_1$ and $m_2$. {\color{black} Note that, the phase modification in Eq.(\ref{delta_Phi}) is a simple propagation effect, which is independent of the  generation effect of GWs. Although this formula is obtained in the stationary phase approximation \cite{Mirshekari:2011yq}, we expect this result is also applicable for the GWs in the more general cases in the presence of spin, precession, eccentricity of compact binaries, and/or for the GWs produced during the merger and ring-down of compact binaries. This extension has been adopted by LIGO and Virgo collaborations in \cite{gw150914,test-gr2}.
}

\subsection{Amplitude modifications}

Now, let us turn to study the effect caused by $\nu_A$. Plugging the decomposition (\ref{decom1}) into (\ref{hAbar}), one finds the equation for $\Phi(t)$,
\bqn\lb{eom_Phi}
i \Phi'' + \Phi'^2 + 2 i \mathcal{H} \Phi' - (1+\mu_A)k^2=0.
\eqn
Similarly, plugging the decomposition (\ref{decom}) and (\ref{decom1}) into (\ref{eom}), one obtains
\bqn \lb{eom2}
&&i (\theta''+\Phi'') + (\Phi'+\theta')^2 \nb\\
&&\;\;\; +i (2+\nu_A)\mathcal{H} (\theta'+\Phi') - (1+\mu_A)k^2=0.
\eqn
Using the equation of motion (\ref{eom_Phi}) for $\Phi$, the above equation reduces to
\bqn
i \theta''+ 2 \theta' \Phi' + \theta'^2 + i (2+\nu_A)\mathcal{H} \theta'+ i \nu_A \mathcal{H} \Phi'=0.\nb\\
\eqn
The phase $\Phi$ is expected to be close to that in GR $\Phi_{\rm GR}$, and $\Phi_{\rm GR}'  \sim k$, where the wave number relates to the GW frequency by $k=2\pi f$. Assuming that
\bqn
\theta'' \ll \Phi'\theta'  \sim k \theta',\;\; k \gg \mathcal{H},
\eqn
and keeping only the leading-order terms, the above equation can be simplified into the form
\bqn
2  \theta' + i \mathcal{H} \nu_A  =0,
\eqn
which has the solution
\bqn
 \theta = -\frac{i}{2} \int_{\tau_e}^{\tau_0} \mathcal{H} \nu_A d\tau.
 \eqn
We observe that the contribution of $\nu_A$ in the phase is purely imaginary. This indicates that the parameter $\nu_A$ leads to modifications of the amplitude of the GWs during the propagation. As a result, relative to the corresponding mode in GR, the amplitude of the left-hand circular polarization of GWs will increase (or decrease) during the propagation, while the amplitude for the right-hand mode will decrease (or increase).

More specifically, one can write the waveform of GWs with parity violation effects in the form
\bqn\lb{waveformA}
h_A = h_{A}^{\rm GR} (1+\delta h_A) e^{ - i \delta \Phi_A},
\eqn
where
\bqn\lb{waveform}
1+ \delta h_A &=&  \exp\left(-\frac{1}{2} \int_{\tau_e}^{\tau_0} \mathcal{H} \nu_A d\tau \right),
\eqn
and $\delta \Phi_A$ is given by (\ref{delta_Phi}). Noticing that
\bqn
\frac{1}{2} \nu_A \mathcal{H} = \frac{1}{2} \left[\ln\left(1- \rho_A \frac{k c_1 }{a\red{M_{\rm PV}}}\right)\right]',
\eqn
we find
\bqn
1+\delta h_A &=& \sqrt{ \frac{1- \rho_A  k c_1(\tau_e) / [a (\tau_e)\red{M_{\rm PV}}] }{1- \rho_A  k c_1(\tau_0) / [a (\tau_0)\red{M_{\rm PV}}] } }\nb\\
&\simeq & 1 + \frac{1}{2} \rho_A k \left(\frac{c_1(\tau_0)}{a(\tau_0) \red{M_{\rm PV}}} - \frac{c_1(\tau_e)}{a(\tau_e) \red{M_{\rm PV}}}\right),
\eqn
which gives
\bqn
\delta h_A & \simeq &  \frac{1}{2} \rho_A k \left(\frac{c_1(\tau_0)}{a(\tau_0) \red{M_{\rm PV}}} - \frac{c_1(\tau_e)}{a(\tau_e) \red{M_{\rm PV}}}\right)\nb\\
&=& \rho_A \frac{ \pi f}{\red{M_{\rm PV}}} \Big[c_1(\tau_0) - (1+z) c_1(\tau_e)\Big].
\eqn
Using $u$ and $\mathcal{M}$, one can rewrite $\delta h_A$ in the form
\bqn\label{delta_h}
\delta h_A = \frac{\rho_{A} u}{ \red{M_{\rm PV}}\mathcal{M}}  \Big[c_1(\tau_0) - (1+z) c_1(\tau_e)\Big].
\eqn
This relation indicates that the amplitude birefringence of GWs depends only on the values of the coefficient $c_1$ at the emitting and observed times.

\subsection{Post-Newtonian orders of the correction terms}

In general, we can write the GWs in the Fourier domain. Similar to the parameterized post-Einsteinian framework of GWs developed in \cite{ppe}, for each circular polarization mode, we can also write the GW waveform as the following parameterized form
\bqn\label{hA-f}
\tilde{h}_A(f) = \tilde{h}_{A}^{\rm GR} (1+ \alpha_{A}^{\rm ppe} u^{a_{A}^{\rm ppe}})e^{i \beta_{A}^{\rm ppe} u^{b_{A}^{\rm ppe}}},
\eqn
where $\alpha_{A}^{\rm ppe} u^{a_{A}^{\rm ppe}}=\delta h_A$ and $\beta_{A}^{\rm ppe} u^{b_{A}^{\rm ppe}}=\delta\Psi_{A}$ represent the amplitude and phase modification respectively. These two terms capture non-GR modifications in the waveform in a generic way. The coefficients $a_A^{\rm ppe}$ and $b_A^{\rm ppe}$ indicate the post-Newtonian (PN) orders of these modifications. In comparison with the results derived in the previous subsection, we obtain that
\bqn
\alpha_{A}^{\rm ppe} &=& \frac{\rho_{A}}{\red{M_{\rm PV}}\mathcal{M}}  \Big[c_1(\tau_0) - (1+z) c_1(\tau_e)\Big], \\
a_{A}^{\rm ppe} &=& 1, \\
\beta_{A}^{\rm ppe} &=&  \xi_A,  \\
b_{A}^{\rm ppe} &=& 2.
\eqn
Let us now count the PN order of these parity violating corrections relative to GR. The relative correction from GR is said to be $n$ PN order if it is proportional to $f^{2n/3}$. Thus, the amplitude correction enters at 1.5 PN order, and the phase correction enters at 5.5 PN order (note that the phase of GR $\propto f^{-5/3}$ at leading order).

\section{Modifications to the GW waveform \label{sec5}}
\renewcommand{\theequation}{5.\arabic{equation}} \setcounter{equation}{0}

In order to make contact with observations, it is convenient to analyze the GWs in the Fourier domain, and the responses of detectors for the GW signals $\tilde{h}(f)$ can be written in terms of waveform of $\tilde{h}_+$ and $\tilde{h}_\times$ as
\bqn
\tilde{h}(f) = [F_+ \tilde{h}_+(f) + F_\times \tilde{h}_\times(f)] e^{- 2 \pi i f \Delta t},
\eqn
where $F_{+}$ and $F_{\times}$ are the beam pattern functions of GW detectors, depending on the source location and polarization angle \cite{F+Fx}. $\Delta t$ is the arrival time difference between the detector and the geocenter. In GR, the waveform of the two polarizations $\tilde{h}_+(f)$ and $\tilde{h}_\times (f)$ are given by
\bqn
\tilde{h}^{\rm GR}_+ = (1+\red{\chi^2}) \mathcal{A}e^{i \Psi}, \;\; \\
\tilde{h}^{\rm GR}_\times = 2 \red{\chi} \mathcal{A} e^{i (\Psi+\pi/2)},
\eqn
where $\mathcal{A}$ and $\Psi$ denote the amplitude and phase of the waveforms $h^{\rm GR}_{+ \; \times}$, and $\chi=\cos\iota$ with $\iota$ being the inclination angle of the binary system. In GR, the explicit forms of $\mathcal{A}$ and $\Psi$ have been calculated in the high-order PN approximation (see for instance \cite{pn} and references therein). Now we would like to derive how the parity violation can affect both the amplitude and the phase of the above waveforms. The circular polarization modes $\tilde{h}_{R}$ and $\tilde{h}_L$ relate to the modes $\tilde{h}_+$ and $\tilde{h}_{\times}$ via
\bqn
\tilde{h}_+ =  \frac{\tilde{h}_{L} + \tilde{h}_{R}}{\sqrt{2}},~~
\tilde{h}_\times =  \frac{\tilde{h}_{L} - \tilde{h}_{R}}{\sqrt{2} i}.
\eqn
Similar to the previous work \cite{Yagi:2017zhb}, throughout this paper, we ignore the parity-violating generation effect, which is caused by a modified energy loss, inspiral rate and chirping rate of the binaries. Since the generation effect occurs on a radiation-reaction time scale, which is much smaller than the GW time of flight, making its impact on the evolution of the GW phase negligible \cite{Alexander:2017jmt}. Thus, the circular polarization modes $\tilde{h}_{A}$ are given in (\ref{hA-f}), and the waveforms for the plus and cross modes become
\bqn
\tilde{h}_+
&\simeq&\tilde{h}^{\rm GR}_+ - (i\delta h-\delta\phi)\tilde{h}^{\rm GR}_\times, \\
\tilde{h}_\times
&\simeq& \tilde{h}_\times^{\rm GR}+(i\delta h-\delta\phi)\tilde{h}_+^{\rm GR},
\eqn
where $\delta\phi \equiv \delta \Psi_{R}$ is given by Eq.(\ref{delta_Phi}) and $\delta h \equiv \delta h_R$ given by Eq.(\ref{delta_h}). Therefore, the Fourier waveform $\tilde{h}(f)$ becomes
\bqn\label{final-hf}
\tilde{h}(f)= \mathcal{A} \delta \mathcal{A} e^{i (\Psi +\delta \Psi )} ,
\eqn
where
\bqn\label{final-delta}
\delta \mathcal{A}&=&\sqrt{(1+\red{\chi^2})^2F^2_+ +4\red{\chi}^2F^2_\times} \nb\\
&&\times\Big[1+\frac{2\red{\chi}(1+\red{\chi}^2)(F^2_+ + F^2_\times)}{(1+\red{\chi}^2)^2F^2_+ +4\red{\chi}^2F^2_\times}\delta h \nb\\
&&-\frac{(1-\red{\chi}^2)^2F_+F_\times}{(1+\red{\chi}^2)^2F^2_+ +4\red{\chi}^2F^2_\times}\delta\phi+\mathcal{O}((\delta h)^2,(\delta\phi)^2)\Big] ,\nb\\
\delta \Psi
&=&\tan^{-1}\left[\frac{2 \red{\chi} F_\times}{(1+\red{\chi}^2)F_+}\right]+\frac{(1-\red{\chi}^2)^2 F_+ F_\times}{(1+\red{\chi}^2)^2 F^2_+ + 4 \red{\chi}^2F^2_\times}\delta h \nb \\
&&+\frac{2 \red{\chi} (1+ \red{\chi}^2)(F^2_+ + F^2_\times)}{(1+ \red{\chi}^2)^2 F^2_+ + 4 \red{\chi}^2F^2_\times}\delta\phi+\mathcal{O}((\delta h)^2,(\delta\phi)^2).\nb\\
\eqn
From these expressions, we find that relative to the waveforms in GR, the modifications of GWs are quantified by the terms $\delta h$ and $\delta \phi$. In the specific case with $\delta h=\delta\phi=0$, the formula in (\ref{final-hf}) returns to that in GR. Since in the parity-violating gravities, the evolution of polarization modes ${h}_+$ and $h_{\times}$ are not independent, the mixture of two modes are inevitable. For this reason, we find that both terms $\delta h$ and $\delta\phi$ appear in the phase and amplitude modifications of $\tilde{h}(f)$. In the CS modified gravity, we have $\delta \phi=0$ and $\delta h\neq 0$, and the formulas in Eq.(\ref{final-delta}) returns to the corresponding ones in \cite{Yagi:2017zhb}. However, in the general ghost-free parity-violating gravities, both correction terms are nonzero. In the leading order, the modification $\delta\mathcal{A}$ (or $\delta\Psi$) linearly depends on $\delta h$ and $\delta \phi$, and it is important to estimate their relative magnitudes. Let us assume the GW is emitted at the redshift $z\sim O(1)$, and approximately treat $c_1$ and $c_2$ as constants during the propagation of GW, we find the ratio of two correction terms is $\delta\phi/\delta h \sim t_0 f$, where $f$ is the GW frequency and $t_0=13.8$ billion years is the cosmic age. As known, $f\sim 100$ Hz for the ground-based GW detectors, and $f\sim 0.01$ Hz for the space-borne detectors. For both cases, we find $\delta\phi$ is more than ten orders of magnitude larger than $\delta h$. So, we arrive at the conclusion: In the general ghost-free parity-violating gravities, both the amplitude and phase corrections of GW waveform $\tilde{h}(f)$ mainly come from the contribution of velocity birefringence rather than that of the amplitude birefringence.

\section{Conclusions \label{sec6}}
\renewcommand{\theequation}{6.\arabic{equation}} \setcounter{equation}{0}

In the parity-violating gravities, the symmetry between left-hand and right-hand circular polarization modes of GWs is broken. So, the effect of birefringence of GWs occurs during their propagation in the universe. In this article, we investigate the GWs in the general ghost-free parity-violating theories of gravity, which is an extension of CS modified gravity. We find that, in general, both amplitude and velocity birefringence effects exist in these theories, which exactly correspond to the amplitude and phase modifications of waveforms for the circular polarization modes. Combining these two modes, we obtain the GW waveforms produced by the compact binary coalescence, and derive the correction terms relative to that in GR. We find that, in the general ghost-free parity-violating theories, the dominant modifications in GW amplitude and phase are both caused mainly by the velocity birefringence effect. Considering the current and potential observations of ground-based and space-borne GW detectors, the explicit waveforms of GWs derived in this article can be used as the template to constrain these theories with parity violation. The comprehensive analysis on this topic will be carried on in a separate paper of this series of works.

\appendix

\section{The coefficients $a_{A}$ and $b_{A}$ in dark energy models}
\renewcommand{\theequation}{A.\arabic{equation}} \setcounter{equation}{0}

In this Appendix, we estimate the values of $c_1$ and $c_2$ in Eqs.(\ref{c1-1}) and (\ref{c2-2}), which depend on the coefficients $\vartheta$, $a_{A}$ and $b_A$, as well as the evolution of the scalar field. Since the scalar field is always motivated to account for the late acceleration of the universe, in this Appendix, we assume that $\phi$-field plays the role of the dark energy, which satisfies the following slow-roll conditions,
\bqn\lb{slowroll}
\dot \phi^2 \ll V(\phi), \;\; |\ddot \phi| \ll  |3 H \dot \phi |.
\eqn
With this condition, we find that the quantities $c_1$ and $c_2$ are slowly varying during the expansion of the universe, which can be approximately treated as constants during low-redshift range. In the expression of $c_1- c_2$, we observe that  it contains only the terms with $a_1, a_2, b_1, b_2, b_3, b_4, b_5, b_7$ and their derivatives with respect to $\phi$. Considering the scalar field $\phi$ with the slow-roll condition (\ref{slowroll}), the leading contribution to $c_1-c_2$ reads
\bqn
\frac{c_1-c_2}{\red{M_{\rm PV}}} &\simeq &8 ( 2 a_1 - a_3 + 2 b_2) M_{\rm Pl}^2 H^3 \epsilon_\phi,
\eqn
where $H$ is the Hubble constant, and $\epsilon_\phi$ is the slow-roll parameter, which is defined as
\bqn
\epsilon_\phi = \frac{\dot \phi^2}{2 M_{\rm Pl}^2 H^2} \ll 1.
\eqn
Note that in the above, we have considered that (i) the coefficients $a_1, \cdots , a_5$, $b_1, \cdots, b_7$, and the derivatives of $a_1$ and $a_3$ with respect to $\phi$ and $X=\phi_\mu \phi^\mu$, i.e., $a_{1,\phi}, a_{1, X}, a_{3,\phi}, a_{3, X}$, are all at the same order of magnitudes, where $a_{A,\phi}\equiv da_A/d\phi$ and $a_{A, X}\equiv da_A/dX$; and (ii) the terms higher than $\mathcal{O}(\epsilon_\phi)$ are ignored in the slow-roll approximation. In the LCDM universe, the magnitudes of $H$ and $\epsilon_\phi$ (determined by the equation-of-state of dark energy) are observables. Thus, the energy scale $\red{M_{\rm PV}}$ of parity violation in the theory is determined by the coefficients $a_A$ and $b_A$. Note that, $c_1$ and $c_2$ can be absorbed by the definition of the energy scale $\red{M_{\rm PV}}$. For a given constraint of $\red{M_{\rm PV}}$ derived from the potential GW observations, we have the following relation to estimate the magnitudes of coefficients $a_A$ and $b_A$,
\bqn
O(a_A,b_A) \sim \frac{1}{8\red{M_{\rm PV}} M_{\rm Pl}^2 H^3 \epsilon_\phi}.
\eqn
Note that, all these coefficients have the unit of ${\rm Energy}^{-6}$.

\section*{Acknowledgements}
We appreciate the helpful discussions with Kai Lin, Xian Gao and Linqing Wen.
J.Q., T.Z., and A.W.  are supported in part by National Natural Science Foundation of China with the Grants No. 11675143, No. 11975203, No. 11675145, the Zhejiang Provincial Natural Science Foundation of China under Grant No. LY20A050002, and the Fundamental Research Funds for the Provincial Universities of Zhejiang in China with Grant No. RF-A2019015. W.Z. is supported by NSFC Grants No. 11773028, No. 11633001, No. 11653002, No. 11421303, No. 11903030, the Fundamental Research Funds for the Central Universities, and the Strategic Priority Research Program of the Chinese Academy of Sciences Grant No. XDB23010200.


\begin{thebibliography}{399}

\bibitem{gw150914}
B. P. Abbottet al.(LIGO Scientific and Virgo Collaborations), Phys. Rev. Lett. {\bf 116}, 061102 (2016); Phys. Rev. D {\bf 93}, 122003 (2016); Phys. Rev. Lett. {\bf 116}, 241102 (2016);116, 131103 (2016).

\bibitem{gw170817}
B. P.  Abbottet  al.(LIGO  Scientific  and  Virgo Collaborations),Phys. Rev. Lett. {\bf 119}, 161101 (2017).

\bibitem{gw-other}
B. P. Abbottet al.(LIGO Scientific and Virgo Collaborations), Phys. Rev. Lett. {\bf 116}, 241103 (2016); Phys. Rev. Lett. {\bf 118}, 221101 (2017); Astrophys. J. {\bf 851}, L35 (2017); Phys. Rev. Lett. {\bf 119}, 141101 (2017); Phys. Rev. X {\bf 9}, 031040 (2019).

\bibitem{test-gr1}
B. P.  Abbottet  al.(LIGO  Scientific  and  Virgo Collaborations), Phys. Rev. Lett. {\bf 123}, 011102 (2019).

\bibitem{test-gr2}
B. P.  Abbottet  al.(LIGO  Scientific  and  Virgo Collaborations), Phys. Rev. D {\bf 100}, 104036 (2019).

\bibitem{test-gr3}
B. P. Abbottet al.(LIGO Scientific and Virgo Collaborations), Phys. Rev. Lett. {\bf 116}, 221101 (2016);

\bibitem{test-gr4}
LIGO Scientific Collaboration, Virgo Collaboration, Fermi Gamma-Ray Burst Monitor, INTEGRAL, Astrophys. J. Lett., {\bf 848}, L13 (2017).

\bibitem{gw-nature}
M. C. Miller and N. Yunes, Nature, {\bf 568}, 496 (2019).

\bibitem{gw-white}
B. S. Sathyaprakash et al., {\emph{Extreme Gravity and Fundamental Physics}}, arXiv:1903.09221.

\bibitem{popper}
K. Popper, {\emph{The Logic of Scientific Discovery}}, 2nd ed.(Routledge Press, London and New York, 2002).


\bibitem{yunes-review}
E. Berti, K. Yagi and N. Yunes, Gen. Relativ. Gravit. {\bf 50}, 46 (2018); E. Berti, K. Yagi, H. Yang and N. Yunes, Gen. Relativ. Gravit. {\bf 50}, 49 (2018); X. Zhang, J. M. Yu, T. Liu, W. Zhao, and A. Z. Wang, Phys. Rev. D {\bf 95}, 124008 (2017);
T. Liu, X. Zhang, W. Zhao, K. Lin, C. Zhang, S. J. Zhang, X. Zhao, T. Zhu, and A. Z.  Wang, Phys. Rev. D {\bf 98}, 083023 (2018).

\bibitem{Ishak:2018his}
 \red{ M.~Ishak,
  Living Rev.\ Rel.\  {\bf 22},  1 (2019).}


\bibitem{Lee-Yang}
T. D. Lee and C. N. Yang, Phys. Rev. {\bf 104}, 254 (1956).

\bibitem{Lue:1998mq}
  A.~Lue, L.~M.~Wang and M.~Kamionkowski,
{Phys.\ Rev.\ Lett.\  {\bf 83}, 1506 (1999)}.


\bibitem{Horava:2009uw}
  P.~Horava,
  {Phys.\ Rev.\ D {\bf 79}, 084008 (2009)}.

\bibitem{Zhu:2011xe}
  T.~Zhu, Q.~Wu, A.~Wang and F.~W.~Shu, {Phys.\ Rev.\ D {\bf 84}, 101502 (2011) (R)}.

  \bibitem{Zhu:2011yu}
  T.~Zhu, F.~W.~Shu, Q.~Wu and A.~Wang,
  {Phys.\ Rev.\ D {\bf 85}, 044053 (2012)}.


  \bibitem{Zhu:2012zk}
  T.~Zhu, Y.~Huang and A.~Wang,
  {JHEP {\bf 1301}, 138 (2013)}; J.~Qiao, G.~H.~Ding, Q.~Wu, T.~Zhu and A.~Wang,
  JCAP {\bf 1909}, 064 (2019).

  \bibitem{wang2013}
  A.~Wang, Q.~Wu, W.~Zhao and T.~Zhu,
  {Phys.\ Rev.\ D {\bf 87}, 103512 (2013)}.

  \bibitem{zhu2013}
  T. Zhu, W. Zhao, Y. Q. Huang, A. Z. Wang and Q. Wu, Phys. Rev. D {\bf 88}, 063508 (2013).


  \bibitem{Wang:2017brl}
  A.~Wang,
  {Int.\ J.\ Mod.\ Phys.\ D {\bf 26}, 1730014 (2017)}.

  \bibitem{Jackiw:2003pm}
  R.~Jackiw and S.~Y.~Pi,
{Phys.\ Rev.\ D {\bf 68}, 104012 (2003)}.

\bibitem{Alexander:2009tp}
  S.~Alexander and N.~Yunes,
{Phys.\ Rept.\  {\bf 480}, 1 (2009)}.

\bibitem{Alexander:2017jmt}
  S.~H.~Alexander and N.~Yunes,
{Phys.\ Rev.\ D {\bf 97}, 064033 (2018)}.



\bibitem{Yunes:2010yf}
  N.~Yunes, R.~O'Shaughnessy, B.~J.~Owen and S.~Alexander,
{Phys.\ Rev.\ D {\bf 82}, 064017 (2010)}.

\bibitem{cs1}
K. Yagi, N. Yunes and T. Tanaka, Phys. Rev. Lett. {\bf 86}, 044037 (2012).


\bibitem{Yagi:2017zhb}
  K.~Yagi and H.~Yang,
{Phys.\ Rev.\ D {\bf 97}, 104018 (2018)}.

\bibitem{Crisostomi:2017ugk}
  M.~Crisostomi, K.~Noui, C.~Charmousis and D.~Langlois,
  {Phys.\ Rev.\ D {\bf 97}, no. 4, 044034 (2018)}.

  \bibitem{Nishizawa:2018srh}
  A.~Nishizawa and T.~Kobayashi,
{Phys.\ Rev.\ D {\bf 98}, 124018 (2018)}.

  \bibitem{Gao:2019liu}
  X.~Gao and X.~Y.~Hong,
  arXiv:1906.07131.


\bibitem{zhao2019}
W. Zhao, T. Liu, L. Q. Wen, T. Zhu, A. Wang, Q. Hu, and C. Zhou, arXiv:1909.13007 [gr-qc].

\bibitem{string}
B. A. Campbell, M. J. Duncan, N. Kalopar and K. A. Olive, Nucl. Phys. B {\bf351}, 778 (1991);
B. A. Campbell, N. Kaloper, R. Madden and K. A. Olive, Nucl. Phys. B {\bf399}, 137 (1993).

\bibitem{CS_E}
 \red{N.~Yunes and D.~N.~Spergel,
  Phys.\ Rev.\ D {\bf 80}, 042004 (2009).}

\bibitem{massive}
S. Deser, R. Jackiw and S. Templeton, Phys. Rev. Lett. {\bf 48}, 975 (1982); Ann. Phys. {\bf 140}, 372 (1982).


\bibitem{spa}
M. Maggiore, {\emph{Theory and Experiments, Gravitational Waves Vol. 1}} (Oxford University Press, Oxford, England, 2007).




\bibitem{Mirshekari:2011yq}
  S.~Mirshekari, N.~Yunes and C.~M.~Will,
{Phys.\ Rev.\ D {\bf 85}, 024041 (2012)}.

\bibitem{ppe}
N. Yunes and F. Pretorius, Phys. Rev. D {\bf 80} 122003 (2009);
N. Cornish, L. Sampson, N. Yunes and F. Pretorius, Phys. Rev. D {\bf 84} 062003 (2011);
K. Chatziioannou, N. Yunes and N. Cornish, Phys. Rev. D {\bf 86} 022004 (2012).

\bibitem{F+Fx}
P. Jaranowski, A. Krolak, and B. F. Schutz, Phys. Rev. D
{\bf 58}, 063001 (1998);
W. Zhao and L. Q. Wen, Phys. Rev. D {\bf 97}, 064031 (2018).

\bibitem{pn}
B. S. Sathyaprakash and B. F. Schutz, Living Rev.
Relativity {\bf 12}, 2 (2009).







\end{thebibliography}
\end{document}